\documentclass[aip,preprint,unsortedaddress]{revtex4-1}

\usepackage{graphicx}
\usepackage{color}
\usepackage{epsfig}
\usepackage{epstopdf}
\usepackage{natbib}
\usepackage{amsmath}
\usepackage{units}

\begin{document}
\title{Magnetoresistance of heavy and light metal/ferromagnet bilayers}
\author{Can Onur Avci}
\author{Kevin Garello}
\author{Johannes Mendil}
\author{Abhijit Ghosh}
\author{Nicolas Blasakis}
\author{Mihai Gabureac}
\author{Morgan Trassin}
\author{Manfred Fiebig}
\author{Pietro Gambardella}
\affiliation{Department of Materials, ETH Z\"{u}rich, H\"{o}nggerbergring 64, CH-8093 Z\"{u}rich, Switzerland}

\date{\today}	
\begin{abstract}	
We studied the magnetoresistance of normal metal (NM)/ferromagnet (FM) bilayers in the linear and nonlinear (current-dependent) regimes and compared it with the amplitude of the spin-orbit torques and thermally induced electric fields. Our experiments reveal that the magnetoresistance of the heavy NM/Co bilayers (NM = Ta, W, Pt) is phenomenologically similar to the spin Hall magnetoresistance (SMR) of YIG/Pt, but has a much larger anisotropy, of the order of $0.5$~\%, which increases with the atomic number of the NM. This SMR-like behavior is absent in light NM/Co bilayers (NM = Ti, Cu), which present the standard AMR expected of polycrystalline FM layers. In the Ta, W, Pt/Co bilayers we find an additional magnetoresistance, directly proportional to the current and to the transverse component of the magnetization. This so-called unidirectional SMR, of the order of 0.005~\%, is largest in W and correlates with the amplitude of the antidamping spin-orbit torque. The unidirectional SMR is below the accuracy of our measurements in YIG/Pt.

\end{abstract}
\maketitle
%
%%%%%%%%%%%%%%%%%%%%%%%%%INTRODUCTION%%%%%%%%%%%%%%%%%%%%%%%%%%%%%%%%%%%%
The interconversion of charge and spin currents is a central theme in spintronics. In normal metal (NM)/ferromagnet (FM) bilayers, the conversion of a charge current into a spin current driven by the spin Hall (SHE)\cite{DyakonovPLA1971} and Rashba-Edelstein effects\cite{EdelsteinSSC1990} leads to strong spin-orbit torques (SOT),\cite{ManchonPRB2008,HaneyPRB2013,AndoPRL2008,MironNM2010,MironN2011,GarelloNN2013,KimNM2013,SanchezNC2013,SkinnerAPL2014} which are widely studied for their role in triggering magnetization switching\cite{MironN2011,LiuS2012,GarelloAPL2014}, magnetic oscillations\cite{LiuPRL2011}, and related applications.\cite{CubukcuAPL2014,DemidovNM2012}
Additionally, it has been shown that the spin currents induced by a charge current can have a significant back-action on the longitudinal charge transport, leading to changes of the resistance of NM/FM bilayers that depend on the relative orientation of the magnetization in the FM and spin-orbit coupling (SOC) induced spin accumulation in the NM.\cite{NakayamaPRL2013,Hahn2013PRB,althammer2013PRB,VlietstraPRB2014,VelezArx2015,AvciNP2015,OlejnikPRB2015}

A direct unequivocal demonstration of such a back-action effect is the spin Hall magnetoresistance (SMR) reported for FM \emph{insulator}/NM bilayers, namely YIG/Pt and YIG/Ta,\cite{NakayamaPRL2013,Hahn2013PRB,althammer2013PRB,VlietstraPRB2014,VelezArx2015} in which complications due to the anisotropic magnetoresistance (AMR) of metallic FM are either absent or restricted to proximity effects in the NM.\cite{Miao2014PRL} For a charge current directed along $x$, the SMR is proportional to $m^{2}_{y}$, the square of the in-plane component of the magnetization transverse to the current, and is typically of the order of 0.01-0.1~\% of the total resistance. In the simplest spin diffusion model, the SMR is associated to the reflection (absorption) of a spin current at the NM/FM interface when the spins are collinear (orthogonal) to the FM magnetization, leading to an increase (decrease) of the conductivity due to the inverse SHE in the NM layer.\cite{NakayamaPRL2013} SMR-like behavior has been observed also in all metal NM/FM systems such as Pt/Co/Pt, Pt/NiFe/Pt, Pt/Co, Ta/Co, and W/CoFeB layers.\cite{KobsPRL2011,*KobsPRB2014,LuPRB2013,AvciNP2015,KimArx2015,ChoSR2015} In this case, however, the SMR cannot be easily singled out due to the AMR of the FM and magnetoresistive contributions induced by spin scattering at the NM/FM interface independent of the SHE.\cite{KobsPRL2011}

\begin{figure}
	\centering	
	\includegraphics[width=15 cm]{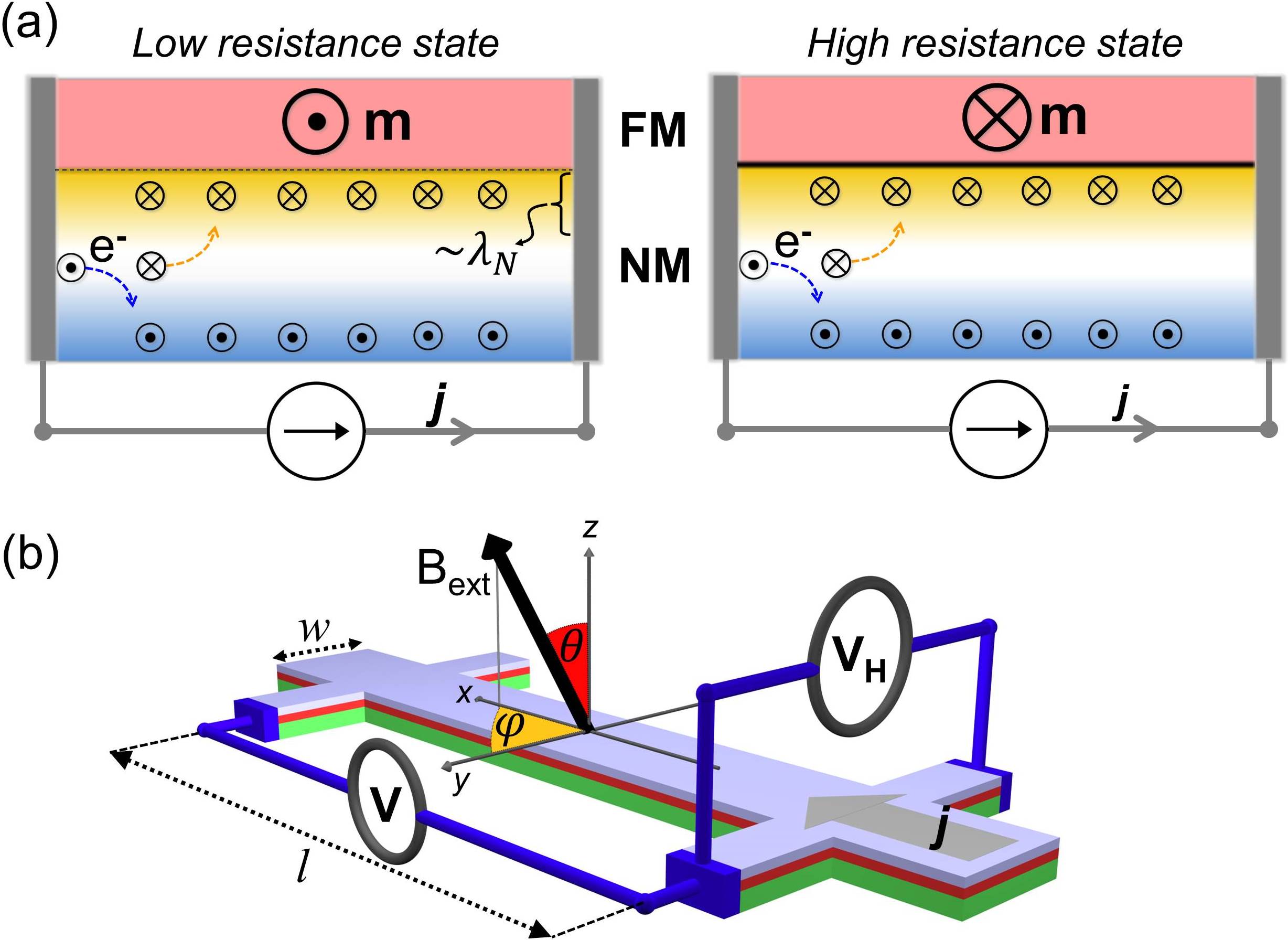}\\	
	\caption{(a) Illustration of the SHE-induced spin accumulation at the NM/FM interface. Parallel (antiparallel) alignment of the magnetization with respect to the spin accumulation gives rise to a decrease (increase) of the longitudinal resistance or USMR. (b) Schematics of the measurement geometry.}\label{fig1}	
\end{figure}

Recently, an additional magnetoresistance has been reported in Pt/Co and Ta/Co bilayers, which depends in magnitude and sign on the product $(\mathbf{j}\times \hat{\mathbf{z}}) \cdot \mathbf{m}$, where $\mathbf{j}$ is the current density and $\mathbf{m}$ the unit vector of the magnetization in the FM.\cite{AvciNP2015} This expression describes a resistance that depends linearly on the applied current and $m_{y}$ (Fig.~\ref{fig1}a), and is therefore a nonlinear effect as opposed to the SMR and AMR, which are both current-independent and proportional to $m^{2}_{y}$ and $m^{2}_{x}$, respectively, as imposed by the Onsager relations in the linear transport regime.\cite{Smith1967} This so-called unidirectional spin Hall magnetoresistance (USMR) is associated to the modulation of the NM/FM interface resistance due to the SHE-induced spin accumulation, similar to the mechanism leading to the current-in-plane giant magnetoresistance in FM/NM/FM multilayers, but orders of magnitude smaller.\cite{AvciNP2015} The USMR depends on the thickness of the NM and is about 0.002-0.003~\% of the total resistance in Ta/Co and Pt/Co for $j=10^7$~A/cm$^2$. An analogous effect has been reported in paramagnetic/ferromagnetic GaMnAs bilayers, where the USMR is significantly larger (0.2~\% for $j=10^6$~A/cm$^2$) due to the much smaller conductivity of semiconductors relative to metals.\cite{OlejnikPRB2015}

These studies reveal that nonlinear phenomena must be taken into account to achieve a full description of the charge-spin conversion in NM/FM systems. New insight may be gained by comparing such effects to the magnetoresistance and SOT, particularly on the nature of the interface resistance, spin accumulation, and material parameters governing them. Further, the USMR offers a way of detecting the magnetization direction of a single FM layer using a two terminal geometry that is otherwise not accessible by conventional magnetoresistance effects. Understanding the role of different NM and FM and searching for systems with larger USMR is a prerequisite to achieve these goals.

Here, we study the magnetoresistance of NM/FM bilayers where the NM has both weak (Ti, Cu) and large (W, Ta, Pt) SOC, as well as low (Cu, Pt) and high (Ti, W, Ta) resistivity. We find that both the SMR-like magnetoresistance and nonlinear USMR are larger in the strong SOC materials, reaching 0.5~\% and 0.004~\% of the total resistance, respectively. The USMR of W/Co is about a factor two (three) larger with respect to Ta/Co (Pt/Co) of equal thickness, in agreement with the larger effective spin Hall angle of W ($\theta_{SH}=0.33\pm 0.05$) estimated from the amplitude of the antidamping SOT in this system. The USMR is found to correlate with the magnitude of the antidamping SOT in the NM/FM layers. Additionally, to separate the USMR from thermomagnetic voltage contributions, we evaluate the electric field due to the anisotropic Nernst (ANE) and spin Seebeck effect (SSE), and show that this correlates with the resistivity of the NM layer. These data are compared to measurements of a YIG/Pt bilayer.

Our samples are NM(6~nm)/Co(2.5~nm)/Al(1.6~nm) layers with NM = Ti, Cu, W, Ta, Pt grown by dc magnetron sputtering on oxidized Si wafers. A 1~nm-thick Ta buffer layer was deposited before the Cu/Co bilayer in order to improve the wetting of the substrate by Cu. The Al capping layer was oxidized by exposure to a radio-frequency O plasma. All samples present isotropic in-plane (easy-plane) magnetization as expected for a polycrystalline Co film. Additionally, a Y$_3$Fe$_5$O$_{12}$ (111) (YIG) (90~nm)/Pt(3~nm) bilayer was grown on a Gd$_3$Ga$_5$O$_{12}$ (111) oriented substrate by a combination of in situ DC sputtering for the metal and pulsed laser deposition for the epitaxial garnet film growth. The crystalline quality and topography of the YIG film was verified using x-ray diffraction and atomic force microscopy, respectively. The as-grown layers were then patterned by optical lithography and ion milling in the form of Hall bars of nominal width $w=4-10~\mu$m and length $l=5w$. The Hall bars were mounted on a motorized stage allowing for in-plane ($\varphi$) and out-of-plane ($\theta$) rotation (see Fig.~\ref{fig1}b), and placed in an electromagnet producing fields of up to 1.7~T. The experiments were performed at room temperature using an ac current of amplitude $j=10^7$~A/cm$^2$ and frequency $\omega/2\pi=10$~Hz. The first and second harmonic resistances, $R_{\omega}$ and $R_{2\omega}$, corresponding to the conventional  (current-independent) resistance and nonlinear (current-dependent) resistance, respectively, and the Hall resistances $R^H_{\omega}$ and $R^H_{2\omega}$ were measured by Fourier analysis of the voltages $V$ and $V_H$ shown in Fig.~\ref{fig1}b (see Ref.~\onlinecite{AvciNP2015} for more details).

\begin{figure}
	\centering	
	\includegraphics[width=15 cm]{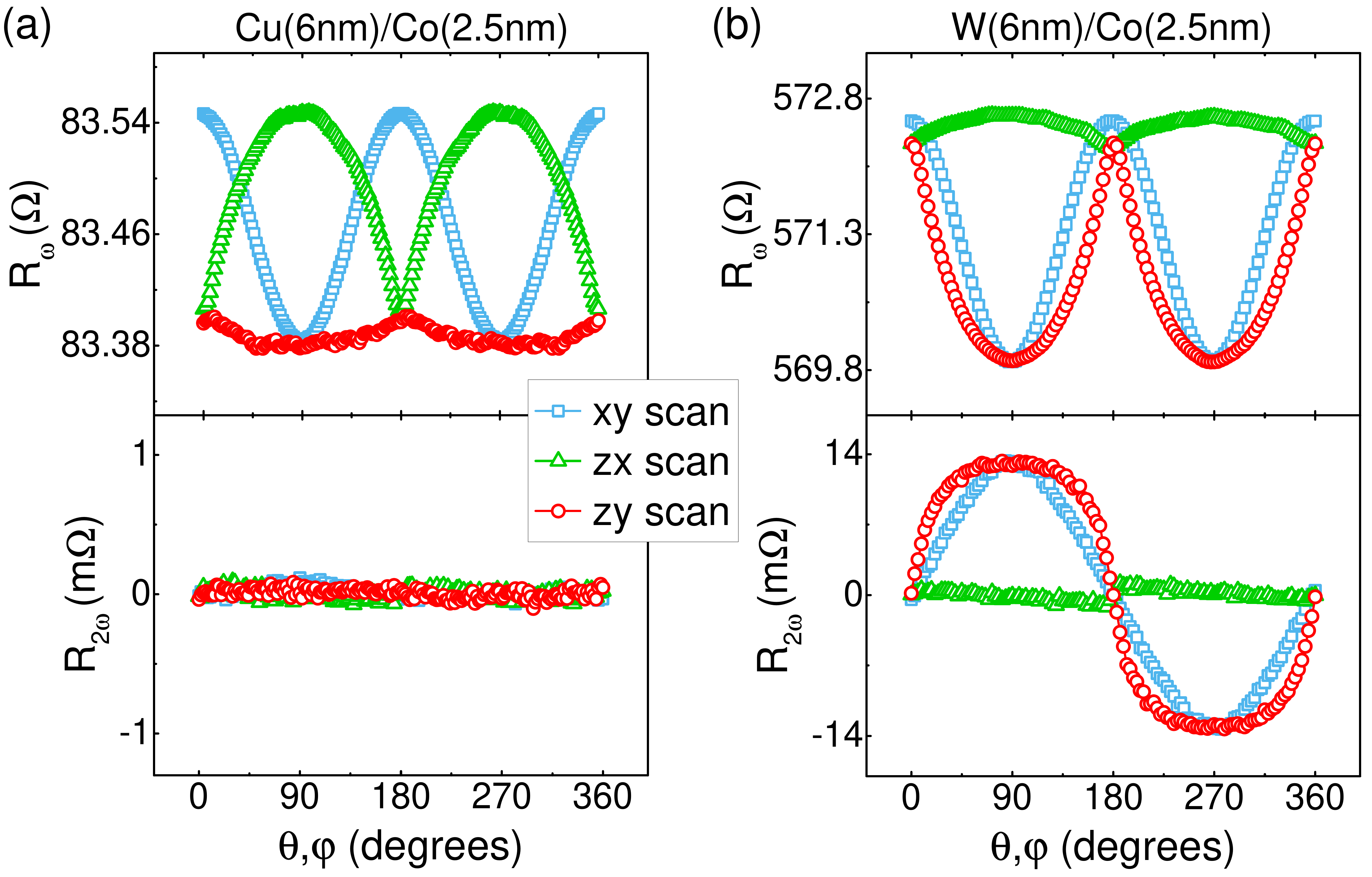}\\
	\caption{Angular dependence of the resistance ($R_{\omega}$, top panels) and nonlinear resistance ($R_{2\omega}$, bottom panels) of (a) Cu/Co and (b) W/Co bilayers with dimensions $w=10$, $l=50$~$\mu$m in an external field of 1.7~T.}\label{fig2}	
\end{figure}

\begin{figure}
	\centering	
	\includegraphics[width=10 cm]{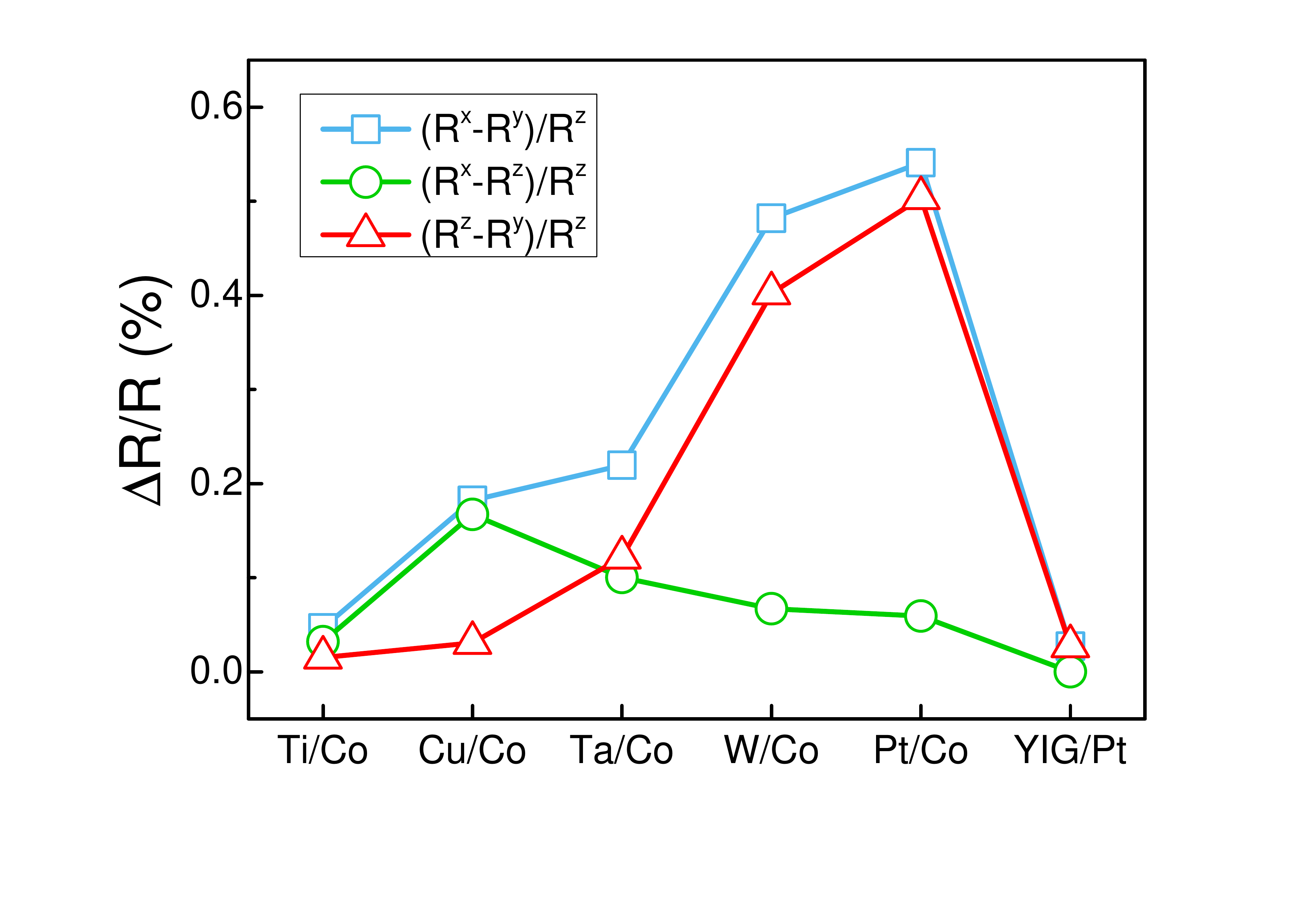}\\	
	\caption{Anisotropy of the magnetoresistance in the $xy$, $zx$, and $zy$ planes derived from the angle-dependent curves shown in Fig.~\ref{fig2}a.}\label{fig3}
\end{figure}

Figure~\ref{fig2} shows the angular dependence of the resistance of (a) Cu/Co and (b) W/Co bilayers measured by sweeping the external field in the $xy$, $zx$, and $zy$ planes defined in Fig.~\ref{fig1}b. The magnetoresistance of the two samples (top panels) is representative of the strong difference between bilayers composed of light and heavy NM: Cu/Co displays the typical AMR of polycrystalline FM layers characterized by $R^x > R^y\approx R^z$, where $R^i$ denotes the resistance measured for $\mathbf{m}$ saturated parallel to $i=x,y,z$, whereas W/Co displays SMR-like behavior, with $R^x \approx R^z > R^y$. The $zy$ magnetoresistance is $(R^z - R^y)/R^z = 0.4$~\% in W/Co, similar to that reported for other metal systems\cite{KobsPRL2011,LuPRB2013,AvciNP2015,KimArx2015} and a factor 15 larger than the SMR of our reference YIG/Pt sample.
Figure~\ref{fig3} resumes the behavior of the different NM/FM bilayers. We find that the $zy$ ($zx$) magnetoresistance increases (decreases) with increasing atomic number of the NM, confirming that the unconventional angular dependence of NM/FM bilayers is related to SOC. The largest $zy$ magnetoresistance is observed for the NM with the largest spin Hall angles, namely W and Pt, consistently with the SMR model.\cite{NakayamaPRL2013} According to this model, the variations of the SMR between the same NM and different FM (as between Pt/Co and Pt/YIG) and between different NM and the same FM (as between W/Co and Pt/Co) may be attributed to changes of the real part of the spin mixing conductance,\cite{NakayamaPRL2013,WeilerPRL2013} which sensitively depends on the material choice and interface properties.\cite{JungfleischAPL2013,IsasaAPL2014} However, these arguments alone are not sufficient to conclude that the $zy$ anisotropy is entirely due to the SMR in all-metal systems, as the anisotropic interface scattering proposed by Kobs et al.\cite{KobsPRL2011,*KobsPRB2014} and other interface contributions\cite{GrigoryanPRB2014} may influence the magnetoresistance.

The second harmonic signals, reported in the lower panel of Fig.~\ref{fig2}b, reveal another striking difference between light and heavy NM systems, namely the presence of a nonlinear resistance in W/Co, which is absent in Cu/Co. We observe that $R_{2\omega}$ has a large variation ($\pm 13.4$~m$\Omega$) in the $xy$ and $zy$ planes and negligibly small variation in the $zx$ plane. $R_{2\omega}$ is found to be proportional to $m_y$ once incomplete saturation of the magnetization in the $zy$ plane is taken into account (due to the competition between the demagnetizing field of Co and  the external field). This signal is compatible with both the USMR and a thermomagnetic contribution due to the anomalous Nernst effect (ANE) for a temperature gradient $\nabla T \parallel z$.\cite{AvciNP2015} In a recent work we have shown that asymmetric heat dissipation towards the air and substrate side gives rise to such an out of plane temperature gradient in NM/FM bilayers, which is more pronounced when the conductivity of the top layer is larger than that of the bottom layer,\cite{AvciPRB2014b} as is the case in W/Co. The ANE signal, however, can be accurately quantified by Hall resistance measurements and separated from the USMR.\cite{AvciNP2015} By saturating the magnetization along $x$ we have quantified the transverse ANE resistance as 0.97~m$\Omega$. Since the ANE is due to an electric field $E_{\nabla T}\propto \nabla T \times \mathbf{m}$, we calculate its longitudinal contribution by using the ratio between the longitudinal and transverse resistance ($\Delta R_{\omega}/\Delta R_{\omega}^H \approx l/w$), which gives 4.09~m$\Omega$, about $30 \%$ of the total $R_{2\omega}$ shown in Fig.~\ref{fig2}b. We thus deduce an USMR value in the W/Co bilayer of $R^{USMR}_{2\omega}=9.3$~m$\Omega$ and the USMR ratio $\Delta R^{USMR}/R = 0.004$~\%, where $\Delta R^{USMR} = R^{USMR}_{2\omega}(+m_y)- R^{USMR}_{2\omega}(-m_y)$. The same procedure was used to estimate the ANE and USMR of all the bilayers studied in this work.

\begin{figure}
	\centering	
	\includegraphics[width=10 cm]{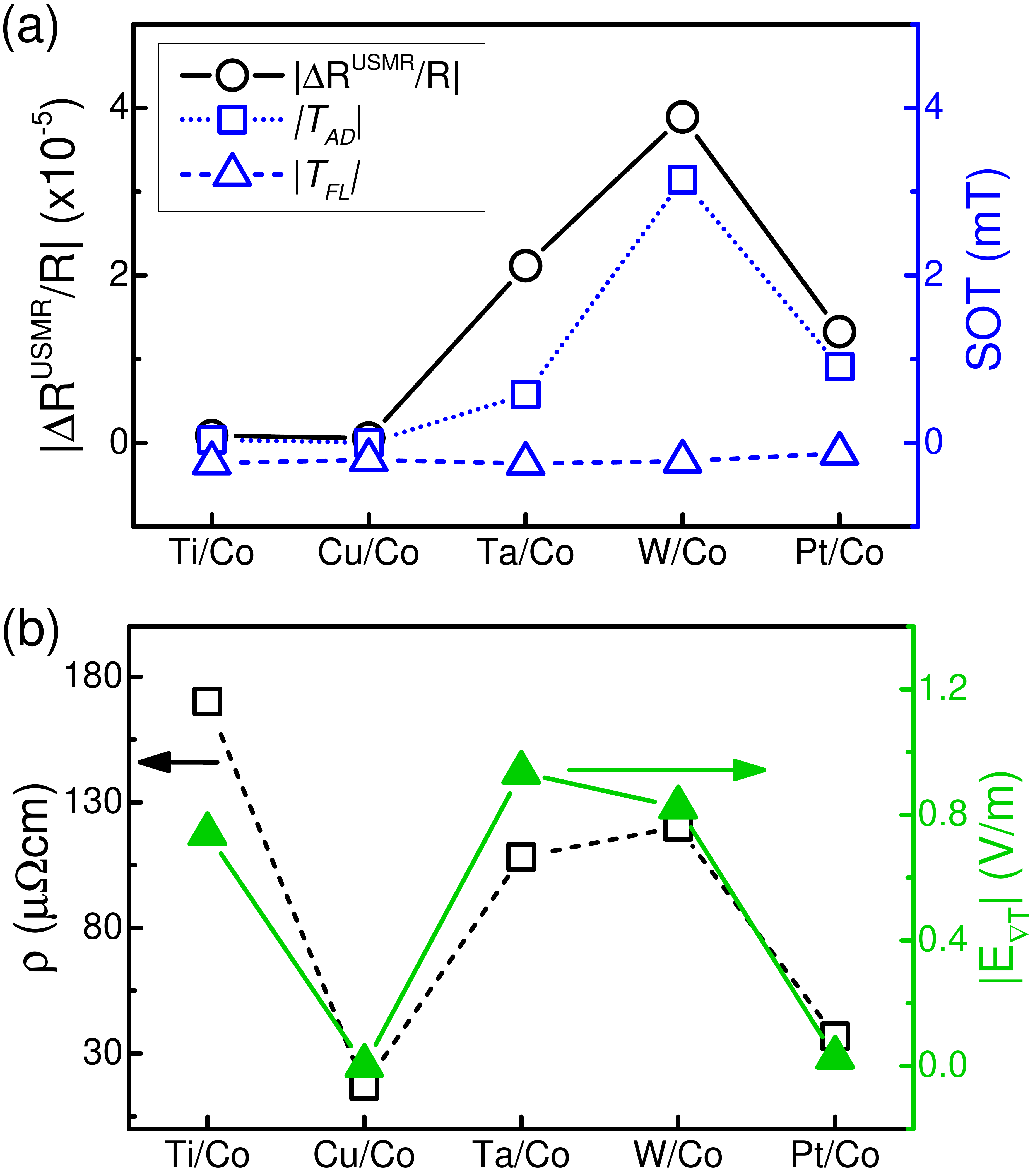}\\	
	\caption{(a) USMR and spin-orbit torques measured in different NM/Co bilayers. (b) Resistivity of the bilayers and electric field field induced by the ANE. The current density is $j=10^7$~A/cm$^2$ in all cases.}\label{fig4}
\end{figure}

The values of $\Delta R^{USMR}/R$ and $E_{\nabla T}$ obtained for different NM are compared in Fig.\ref{fig4}a and b. We find that the USMR is about a factor two (three) larger in W/Co with respect to Ta/Co (Pt/Co), and has opposite sign in Ta/Co and W/Co relative to Pt/Co. Although the amplitude of $\Delta R^{USMR}/R$ depends on the ratio between the thickness of the NM, $t_{NM}$, and the spin diffusion length, $\lambda_{NM}$,\cite{AvciNP2015} the similar $\lambda_{NM}$ of Ta, W, and Pt indicates that the USMR is strongly enhanced in W, which we associate to the larger spin Hall angle of $\beta$-phase W relative to Ta and Pt.\cite{PaiAPL2012} Contrary to the USMR, we find that the ANE scales with the resistivity of the layers independently of the SOC in the NM. The ANE-induced electric field is of the order of 1~V/m for Ti, Ta, and W, all of them highly resistive metals with $\rho$ exceeding 100~$\mu\Omega\cdot$cm, whereas negligible ANE signals are detected in NM with low resistivity, where the current is shunted towards the NM side of bilayer. This indicates that the dominant ANE contribution comes from "bulk" Co and is largest when the current flows through the top FM layer.

As noted in the introduction, the spin currents induced by charge flow are responsible for the USMR as well as SOT. In the following, we compare the magnitude of these effects in different layers. The antidamping and field-like SOT, $T_{AD}$ and $T_{FL}$, were measured using harmonic Hall voltage analysis,\cite{GarelloNN2013,KimNM2013} carried out simultaneously with the resistance measurements. The details of such measurements in samples with in-plane magnetization are outlined in Ref.~\onlinecite{AvciPRB2014b}. Figure~\ref{fig4}a evidences a clear correlation between $T_{AD}$ and $\Delta R^{USMR}/R$ in W/Co and Pt/Co, and, to a lesser degree, in Ta/Co. Both quantities have negligible amplitude in the light NM, as expected due to the small SOC of these systems. Assuming that $T_{AD}$ is driven by the SHE of the NM and a transparent interface, the torque amplitude can be expressed as an effective spin Hall angle\cite{LiuPRL2011} $\theta_{SH}=\frac{2e}{\hbar}\frac{M_{s} t_{FM}}{j}[1-\mathrm{sech}(\frac{t_{NM}}{\lambda_{NM}})]T_{AD}$, where $\mu_0 M_s=1.5$~T is the saturation magnetization of Co, $t_{NM} =6$~nm, $\lambda_{W}=1.6$~nm, $\lambda_{Pt}=1.1$~nm, and $\lambda_{Ta}=1.5$~nm.\cite{AvciNP2015,KimArx2015} We thus obtain $\theta_{SH}$(W)$= 0.33 \pm 0.05$, $\theta_{SH}$(Pt)$=0.10 \pm 0.02$, and $\theta_{SH}$(Ta)$=0.09 \pm 0.02$, comparable to previous reports.\cite{PaiAPL2012,GarelloNN2013,LiuS2012,AvciPRB2014}

The correspondence between the USMR and $\theta_{SH}$ of W shows that materials with large spin Hall angles are required to enhance this effect. The fact that $\Delta R^{USMR}/R$ of Pt is smaller than Ta whereas $\theta_{SH}$(Pt)$>\theta_{SH}$(Ta), on the other hand, is attributed to the dilution of the USMR in highly conducting NM layers when $t_{NM} > \lambda_{NM}$,\cite{AvciNP2015} as is the case here for $t_{NM} =6$~nm, although other effects may also play a role, such as spin memory loss at the NM/FM interface.\cite{bass2014jmmm} Additionally, we observe that $T_{FL}$ is much smaller than $T_{AD}$ in the heavy NM, as expected when the thickness of the FM exceeds $\sim 1$~nm,\cite{KimNM2013, AvciPRB2014b} and has no apparent relationship to the USMR. The latter observation suggests that the USMR depends mainly on the real part of the spin mixing conductance, which is proportional to $T_{AD}$, similar to the SMR.\cite{NakayamaPRL2013}

\begin{figure}
	\centering	
	\includegraphics[width=15 cm]{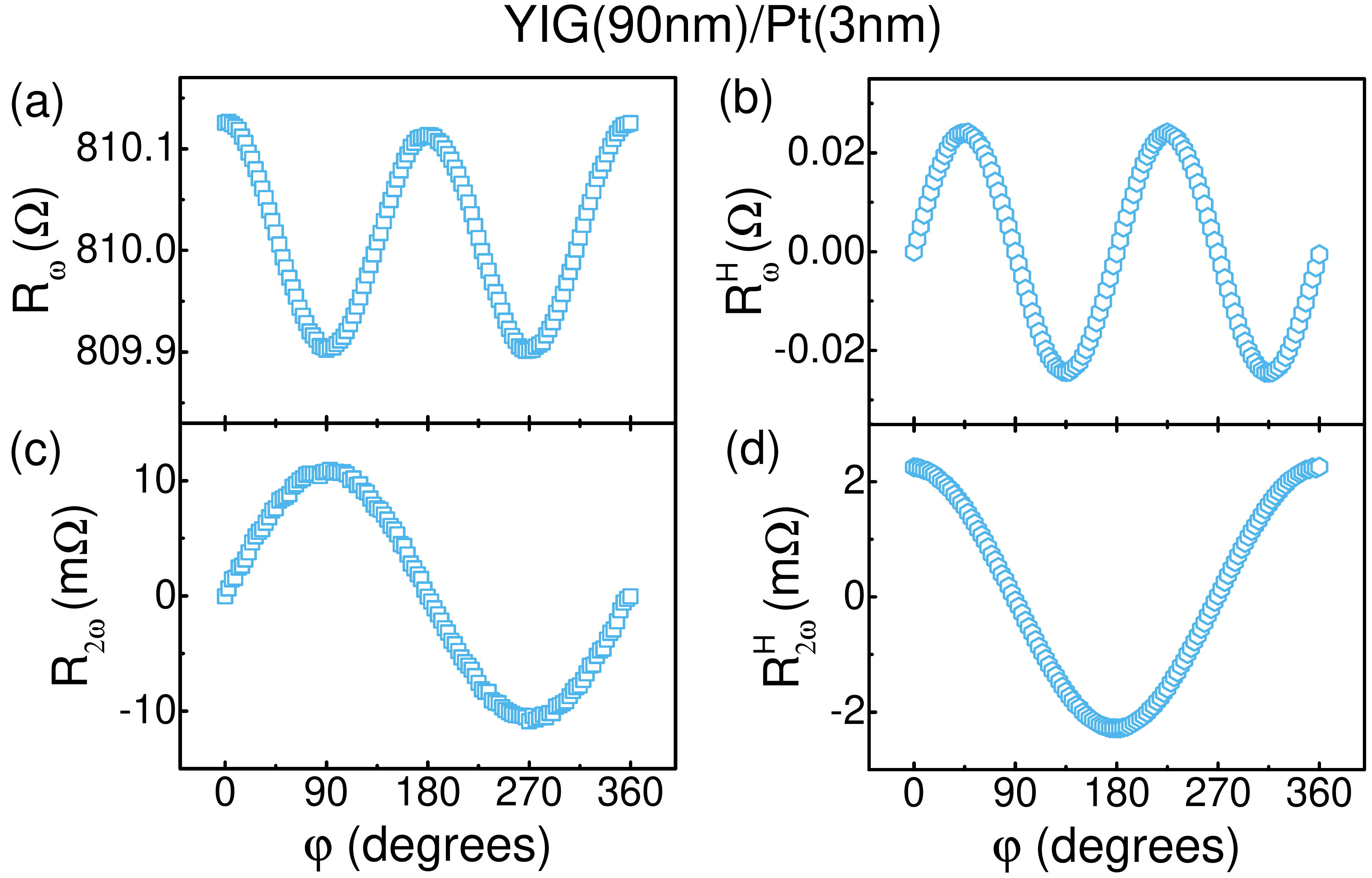}\\	
	\caption{Angular dependence of the longitudinal resistance (a) and transverse resistance (b) of YIG(90~nm)/Pt(3~nm) with dimensions $w=10$, $l=50$~$\mu$m measured in an external field of 0.2~T. Nonlinear longitudinal resistance (c) and transverse resistance (d). The curve in (c) is an average over twenty consecutive measurements to improve the signal-to-noise ratio.}\label{fig5}
\end{figure}

Finally, we show that the USMR is absent when the FM is an insulator such as YIG, as expected by analogy to the current-in-plane giant magnetoresistance. We use YIG/Pt as a model FM insulator/NM system with well-characterized SMR and thermomagnetic properties,\cite{NakayamaPRL2013,Hahn2013PRB,althammer2013PRB,VlietstraPRB2014} and show that no significant USMR signal can be detected in this system. Figures~\ref{fig5}a and b show the angular dependence of the longitudinal ($R_{\omega}$) and transverse ($R_{\omega}^H$) resistance of a YIG(90~nm)/Pt(3~nm) bilayer in the $xy$ plane. In both channels we measure a signal consistent with the SMR, namely $R_{\omega} \sim (1-m_{y}^{2}) = \cos^2\varphi$ and $R_{\omega}^H \sim m_{x}m_{y}=\sin2\varphi$, with ratio $\Delta R_{\omega}/\Delta R_{\omega}^H = 4.3 \approx l/w $ as determined from optical microscopy. From the longitudinal measurement we calculate the resistivity of Pt, $\rho_{Pt}=56.5$~$\mu\Omega$cm, and the SMR ratio $\frac{R^{x,z}-R^{y}}{R^{x,z}}=2.7 \cdot 10^{-4}$, both within the range of literature values reported for samples with comparable Pt and YIG thickness.\cite{althammer2013PRB,vlietstra2013PRB}

The second harmonic resistances, $R_{2\omega}$ and $R_{2\omega}^H$, are shown in Fig.~\ref{fig5}c and d. The angular variation of $R_{2\omega}$ in the $xy$ plane is about a factor ten smaller relative to $R_{\omega}$ and is proportional to $m_y=\sin\varphi$. This signal has the symmetry expected of the USMR as well as the spin Seebeck effect (SSE) due to an out of plane thermal gradient.\cite{SchreierAPL2013,VlietstraPRB2014} Similar to the ANE in FM metals, the SSE voltage appears in the longitudinal channel when $\mathbf{m} \parallel y$ and in the transverse channel when $\mathbf{m} \parallel x$. Accordingly, we observe that $R_{2\omega}^H$ is proportional to $m_x=\cos\varphi$ in Fig.~\ref{fig5}d and estimate the electric field due to the SSE as $E_{\nabla T} = 0.68$~V/m. We can thus calculate the thermal contribution to $R_{2\omega}$ by rescaling the transverse resistance $R_{2\omega}^H$ by the factor $\Delta R_{\omega}/\Delta R_{\omega}^H$, as done in the case of the NM/FM metal layers. The comparison between $R_{2\omega}$ and $R_{2\omega}^H$ in Fig.~\ref{fig5} shows that $\Delta R_{2\omega}/\Delta R_{2\omega}^H \approx \Delta R_{\omega}/\Delta R_{\omega}^H$ to within 10~\% accuracy. Therefore, we conclude that most of the $R_{2\omega}$ signal is of thermal origin and not related to the USMR. The small discrepancy between the longitudinal and rescaled transverse nonlinear signals can be explained by several factors, for example by considering that the current spreading in the Hall branches can decrease Joule heating in the Hall cross with respect to the central region of the Hall bar, thus reducing the thermal voltage in the transverse measurement with respect to the longitudinal one. Alternatively, a small proximity-induced magnetization in Pt could couple to the spin accumulation due to the SHE and give rise to the USMR. Overall, our data show that the USMR in YIG/Pt, if it exists, is much smaller compared to NM/FM metal systems.\\
\indent In summary, we have measured the angular dependence of the magnetoresistance in light and heavy metal/FM layers in the linear and nonlinear response regimes. The resistance of Pt/Co, W/Co, and Ta/Co bilayers depends strongly on the magnetization orientation in the plane perpendicular to the current direction, akin to the spin Hall magnetoresistance (SMR) in YIG/Pt, but with magnetoresistance ratios 15 times as large, of the order of $(R^{z} - R^y)/R^{z} = 0.5$~\%. This ratio increases with the atomic number of the NM, whereas the light NM/Co bilayers (NM = Ti, Cu) present the usual AMR expected of polycrystalline FM layers, characterized by $R^z \approx R^y$. Thermomagnetic effects typified by the ANE correlate with the resistivity of the NM rather than SOC. In the Ta, W, Pt/Co bilayers we find an additional \emph{nonlinear} magnetoresistance, which depends \emph{linearly} on the current and on the $y$-component of the magnetization. This so-called USMR, of the order of 0.005~\%, is enhanced by a factor 2-3 in W/Co relative to Pt/Co and Ta/Co and correlates with the amplitude of the AD spin-orbit torque, whereas it shows no apparent relationship to the FL spin-orbit torque. The USMR is below the accuracy of our measurements in YIG/Pt. These results suggest that NM with large spin Hall angles and NM/FM interfaces with large and real spin mixing conductance are required to enhance the USMR.\\

\begin{acknowledgments}
This research was supported by the Swiss National Science Foundation (Grant No. 200021-153404) and the European Commission under the Seventh Framework Program (spOt project, Grant No. 318144).
\end{acknowledgments}

\bibliographystyle{aipnum4-1}

\begin{thebibliography}{40}%
\makeatletter
\providecommand \@ifxundefined [1]{%
 \@ifx{#1\undefined}
}%
\providecommand \@ifnum [1]{%
 \ifnum #1\expandafter \@firstoftwo
 \else \expandafter \@secondoftwo
 \fi
}%
\providecommand \@ifx [1]{%
 \ifx #1\expandafter \@firstoftwo
 \else \expandafter \@secondoftwo
 \fi
}%
\providecommand \natexlab [1]{#1}%
\providecommand \enquote  [1]{``#1''}%
\providecommand \bibnamefont  [1]{#1}%
\providecommand \bibfnamefont [1]{#1}%
\providecommand \citenamefont [1]{#1}%
\providecommand \href@noop [0]{\@secondoftwo}%
\providecommand \href [0]{\begingroup \@sanitize@url \@href}%
\providecommand \@href[1]{\@@startlink{#1}\@@href}%
\providecommand \@@href[1]{\endgroup#1\@@endlink}%
\providecommand \@sanitize@url [0]{\catcode `\\12\catcode `\$12\catcode
  `\&12\catcode `\#12\catcode `\^12\catcode `\_12\catcode `\%12\relax}%
\providecommand \@@startlink[1]{}%
\providecommand \@@endlink[0]{}%
\providecommand \url  [0]{\begingroup\@sanitize@url \@url }%
\providecommand \@url [1]{\endgroup\@href {#1}{\urlprefix }}%
\providecommand \urlprefix  [0]{URL }%
\providecommand \Eprint [0]{\href }%
\providecommand \doibase [0]{http://dx.doi.org/}%
\providecommand \selectlanguage [0]{\@gobble}%
\providecommand \bibinfo  [0]{\@secondoftwo}%
\providecommand \bibfield  [0]{\@secondoftwo}%
\providecommand \translation [1]{[#1]}%
\providecommand \BibitemOpen [0]{}%
\providecommand \bibitemStop [0]{}%
\providecommand \bibitemNoStop [0]{.\EOS\space}%
\providecommand \EOS [0]{\spacefactor3000\relax}%
\providecommand \BibitemShut  [1]{\csname bibitem#1\endcsname}%
\let\auto@bib@innerbib\@empty
%</preamble>
\bibitem [{\citenamefont {Dyakonov}\ and\ \citenamefont
  {Perel}(1971)}]{DyakonovPLA1971}%
  \BibitemOpen
  \bibfield  {author} {\bibinfo {author} {\bibfnamefont {M.}~\bibnamefont
  {Dyakonov}}\ and\ \bibinfo {author} {\bibfnamefont {V.}~\bibnamefont
  {Perel}},\ }\href@noop {} {\bibfield  {journal} {\bibinfo  {journal} {Phys.
  Lett. A}\ }\textbf {\bibinfo {volume} {35}},\ \bibinfo {pages} {459}
  (\bibinfo {year} {1971})}\BibitemShut {NoStop}%
\bibitem [{\citenamefont {Edelstein}(1990)}]{EdelsteinSSC1990}%
  \BibitemOpen
  \bibfield  {author} {\bibinfo {author} {\bibfnamefont {V.~M.}\ \bibnamefont
  {Edelstein}},\ }\href@noop {} {\bibfield  {journal} {\bibinfo  {journal}
  {Sol. St. Comm.}\ }\textbf {\bibinfo {volume} {73}},\ \bibinfo {pages} {233}
  (\bibinfo {year} {1990})}\BibitemShut {NoStop}%
\bibitem [{\citenamefont {Manchon}\ and\ \citenamefont
  {Zhang}(2008)}]{ManchonPRB2008}%
  \BibitemOpen
  \bibfield  {author} {\bibinfo {author} {\bibfnamefont {A.}~\bibnamefont
  {Manchon}}\ and\ \bibinfo {author} {\bibfnamefont {S.}~\bibnamefont
  {Zhang}},\ }\href@noop {} {\bibfield  {journal} {\bibinfo  {journal} {Phys.
  Rev. B}\ }\textbf {\bibinfo {volume} {78}},\ \bibinfo {pages} {212405}
  (\bibinfo {year} {2008})}\BibitemShut {NoStop}%
\bibitem [{\citenamefont {Haney}\ \emph {et~al.}(2013)\citenamefont {Haney},
  \citenamefont {Lee}, \citenamefont {Lee}, \citenamefont {Manchon},\ and\
  \citenamefont {Stiles}}]{HaneyPRB2013}%
  \BibitemOpen
  \bibfield  {author} {\bibinfo {author} {\bibfnamefont {P.~M.}\ \bibnamefont
  {Haney}}, \bibinfo {author} {\bibfnamefont {H.-W.}\ \bibnamefont {Lee}},
  \bibinfo {author} {\bibfnamefont {K.-J.}\ \bibnamefont {Lee}}, \bibinfo
  {author} {\bibfnamefont {A.}~\bibnamefont {Manchon}}, \ and\ \bibinfo
  {author} {\bibfnamefont {M.}~\bibnamefont {Stiles}},\ }\href@noop {}
  {\bibfield  {journal} {\bibinfo  {journal} {Phys. Rev. B}\ }\textbf {\bibinfo
  {volume} {87}},\ \bibinfo {pages} {174411} (\bibinfo {year}
  {2013})}\BibitemShut {NoStop}%
\bibitem [{\citenamefont {Ando}\ \emph {et~al.}(2008)\citenamefont {Ando},
  \citenamefont {Takahashi}, \citenamefont {Harii}, \citenamefont {Sasage},
  \citenamefont {Ieda}, \citenamefont {Maekawa},\ and\ \citenamefont
  {Saitoh}}]{AndoPRL2008}%
  \BibitemOpen
  \bibfield  {author} {\bibinfo {author} {\bibfnamefont {K.}~\bibnamefont
  {Ando}}, \bibinfo {author} {\bibfnamefont {S.}~\bibnamefont {Takahashi}},
  \bibinfo {author} {\bibfnamefont {K.}~\bibnamefont {Harii}}, \bibinfo
  {author} {\bibfnamefont {K.}~\bibnamefont {Sasage}}, \bibinfo {author}
  {\bibfnamefont {J.}~\bibnamefont {Ieda}}, \bibinfo {author} {\bibfnamefont
  {S.}~\bibnamefont {Maekawa}}, \ and\ \bibinfo {author} {\bibfnamefont
  {E.}~\bibnamefont {Saitoh}},\ }\href@noop {} {\bibfield  {journal} {\bibinfo
  {journal} {Phys. Rev. Lett.}\ }\textbf {\bibinfo {volume} {101}},\ \bibinfo
  {pages} {036601} (\bibinfo {year} {2008})}\BibitemShut {NoStop}%
\bibitem [{\citenamefont {Miron}\ \emph {et~al.}(2010)\citenamefont {Miron},
  \citenamefont {Gaudin}, \citenamefont {Auffret}, \citenamefont {Rodmacq},
  \citenamefont {Schuhl}, \citenamefont {Pizzini}, \citenamefont {Vogel},\ and\
  \citenamefont {Gambardella}}]{MironNM2010}%
  \BibitemOpen
  \bibfield  {author} {\bibinfo {author} {\bibfnamefont {I.~M.}\ \bibnamefont
  {Miron}}, \bibinfo {author} {\bibfnamefont {G.}~\bibnamefont {Gaudin}},
  \bibinfo {author} {\bibfnamefont {S.}~\bibnamefont {Auffret}}, \bibinfo
  {author} {\bibfnamefont {B.}~\bibnamefont {Rodmacq}}, \bibinfo {author}
  {\bibfnamefont {A.}~\bibnamefont {Schuhl}}, \bibinfo {author} {\bibfnamefont
  {S.}~\bibnamefont {Pizzini}}, \bibinfo {author} {\bibfnamefont
  {J.}~\bibnamefont {Vogel}}, \ and\ \bibinfo {author} {\bibfnamefont
  {P.}~\bibnamefont {Gambardella}},\ }\href@noop {} {\bibfield  {journal}
  {\bibinfo  {journal} {Nature Mater.}\ }\textbf {\bibinfo {volume} {9}},\
  \bibinfo {pages} {230} (\bibinfo {year} {2010})}\BibitemShut {NoStop}%
\bibitem [{\citenamefont {Miron}\ \emph {et~al.}(2011)\citenamefont {Miron},
  \citenamefont {Garello}, \citenamefont {Gaudin}, \citenamefont {Zermatten},
  \citenamefont {Costache}, \citenamefont {Auffret}, \citenamefont {Bandiera},
  \citenamefont {Rodmacq}, \citenamefont {Schuhl},\ and\ \citenamefont
  {Gambardella}}]{MironN2011}%
  \BibitemOpen
  \bibfield  {author} {\bibinfo {author} {\bibfnamefont {I.~M.}\ \bibnamefont
  {Miron}}, \bibinfo {author} {\bibfnamefont {K.}~\bibnamefont {Garello}},
  \bibinfo {author} {\bibfnamefont {G.}~\bibnamefont {Gaudin}}, \bibinfo
  {author} {\bibfnamefont {P.-J.}\ \bibnamefont {Zermatten}}, \bibinfo {author}
  {\bibfnamefont {M.~V.}\ \bibnamefont {Costache}}, \bibinfo {author}
  {\bibfnamefont {S.}~\bibnamefont {Auffret}}, \bibinfo {author} {\bibfnamefont
  {S.}~\bibnamefont {Bandiera}}, \bibinfo {author} {\bibfnamefont
  {B.}~\bibnamefont {Rodmacq}}, \bibinfo {author} {\bibfnamefont
  {A.}~\bibnamefont {Schuhl}}, \ and\ \bibinfo {author} {\bibfnamefont
  {P.}~\bibnamefont {Gambardella}},\ }\href@noop {} {\bibfield  {journal}
  {\bibinfo  {journal} {Nature}\ }\textbf {\bibinfo {volume} {476}},\ \bibinfo
  {pages} {189} (\bibinfo {year} {2011})}\BibitemShut {NoStop}%
\bibitem [{\citenamefont {Garello}\ \emph {et~al.}(2013)\citenamefont
  {Garello}, \citenamefont {Miron}, \citenamefont {Avci}, \citenamefont
  {Freimuth}, \citenamefont {Mokrousov}, \citenamefont {Bl{\"u}gel},
  \citenamefont {Auffret}, \citenamefont {Boulle}, \citenamefont {Gaudin},\
  and\ \citenamefont {Gambardella}}]{GarelloNN2013}%
  \BibitemOpen
  \bibfield  {author} {\bibinfo {author} {\bibfnamefont {K.}~\bibnamefont
  {Garello}}, \bibinfo {author} {\bibfnamefont {I.~M.}\ \bibnamefont {Miron}},
  \bibinfo {author} {\bibfnamefont {C.~O.}\ \bibnamefont {Avci}}, \bibinfo
  {author} {\bibfnamefont {F.}~\bibnamefont {Freimuth}}, \bibinfo {author}
  {\bibfnamefont {Y.}~\bibnamefont {Mokrousov}}, \bibinfo {author}
  {\bibfnamefont {S.}~\bibnamefont {Bl{\"u}gel}}, \bibinfo {author}
  {\bibfnamefont {S.}~\bibnamefont {Auffret}}, \bibinfo {author} {\bibfnamefont
  {O.}~\bibnamefont {Boulle}}, \bibinfo {author} {\bibfnamefont
  {G.}~\bibnamefont {Gaudin}}, \ and\ \bibinfo {author} {\bibfnamefont
  {P.}~\bibnamefont {Gambardella}},\ }\href@noop {} {\bibfield  {journal}
  {\bibinfo  {journal} {Nature Nanotech.}\ }\textbf {\bibinfo {volume} {8}},\
  \bibinfo {pages} {587} (\bibinfo {year} {2013})}\BibitemShut {NoStop}%
\bibitem [{\citenamefont {Kim}\ \emph {et~al.}(2013)\citenamefont {Kim},
  \citenamefont {Sinha}, \citenamefont {Hayashi}, \citenamefont {Yamanouchi},
  \citenamefont {Fukami}, \citenamefont {Suzuki}, \citenamefont {Mitani},\ and\
  \citenamefont {Ohno}}]{KimNM2013}%
  \BibitemOpen
  \bibfield  {author} {\bibinfo {author} {\bibfnamefont {J.}~\bibnamefont
  {Kim}}, \bibinfo {author} {\bibfnamefont {J.}~\bibnamefont {Sinha}}, \bibinfo
  {author} {\bibfnamefont {M.}~\bibnamefont {Hayashi}}, \bibinfo {author}
  {\bibfnamefont {M.}~\bibnamefont {Yamanouchi}}, \bibinfo {author}
  {\bibfnamefont {S.}~\bibnamefont {Fukami}}, \bibinfo {author} {\bibfnamefont
  {T.}~\bibnamefont {Suzuki}}, \bibinfo {author} {\bibfnamefont
  {S.}~\bibnamefont {Mitani}}, \ and\ \bibinfo {author} {\bibfnamefont
  {H.}~\bibnamefont {Ohno}},\ }\href@noop {} {\bibfield  {journal} {\bibinfo
  {journal} {Nature Mater.}\ }\textbf {\bibinfo {volume} {12}},\ \bibinfo
  {pages} {240} (\bibinfo {year} {2013})}\BibitemShut {NoStop}%
\bibitem [{\citenamefont {S{\'a}nchez}\ \emph {et~al.}(2013)\citenamefont
  {S{\'a}nchez}, \citenamefont {Vila}, \citenamefont {Desfonds}, \citenamefont
  {Gambarelli}, \citenamefont {Attan{\'e}}, \citenamefont {De~Teresa},
  \citenamefont {Mag{\'e}n},\ and\ \citenamefont {Fert}}]{SanchezNC2013}%
  \BibitemOpen
  \bibfield  {author} {\bibinfo {author} {\bibfnamefont {J.~R.}\ \bibnamefont
  {S{\'a}nchez}}, \bibinfo {author} {\bibfnamefont {L.}~\bibnamefont {Vila}},
  \bibinfo {author} {\bibfnamefont {G.}~\bibnamefont {Desfonds}}, \bibinfo
  {author} {\bibfnamefont {S.}~\bibnamefont {Gambarelli}}, \bibinfo {author}
  {\bibfnamefont {J.}~\bibnamefont {Attan{\'e}}}, \bibinfo {author}
  {\bibfnamefont {J.}~\bibnamefont {De~Teresa}}, \bibinfo {author}
  {\bibfnamefont {C.}~\bibnamefont {Mag{\'e}n}}, \ and\ \bibinfo {author}
  {\bibfnamefont {A.}~\bibnamefont {Fert}},\ }\href@noop {} {\bibfield
  {journal} {\bibinfo  {journal} {Nat. Comm.}\ }\textbf {\bibinfo {volume} {4}}
  (\bibinfo {year} {2013})}\BibitemShut {NoStop}%
\bibitem [{\citenamefont {Skinner}\ \emph {et~al.}(2014)\citenamefont
  {Skinner}, \citenamefont {Wang}, \citenamefont {Hindmarch}, \citenamefont
  {Rushforth}, \citenamefont {Irvine}, \citenamefont {Heiss}, \citenamefont
  {Kurebayashi},\ and\ \citenamefont {Ferguson}}]{SkinnerAPL2014}%
  \BibitemOpen
  \bibfield  {author} {\bibinfo {author} {\bibfnamefont {T.}~\bibnamefont
  {Skinner}}, \bibinfo {author} {\bibfnamefont {M.}~\bibnamefont {Wang}},
  \bibinfo {author} {\bibfnamefont {A.}~\bibnamefont {Hindmarch}}, \bibinfo
  {author} {\bibfnamefont {A.}~\bibnamefont {Rushforth}}, \bibinfo {author}
  {\bibfnamefont {A.}~\bibnamefont {Irvine}}, \bibinfo {author} {\bibfnamefont
  {D.}~\bibnamefont {Heiss}}, \bibinfo {author} {\bibfnamefont
  {H.}~\bibnamefont {Kurebayashi}}, \ and\ \bibinfo {author} {\bibfnamefont
  {A.}~\bibnamefont {Ferguson}},\ }\href@noop {} {\bibfield  {journal}
  {\bibinfo  {journal} {Applied Physics Letters}\ }\textbf {\bibinfo {volume}
  {104}},\ \bibinfo {pages} {062401} (\bibinfo {year} {2014})}\BibitemShut
  {NoStop}%
\bibitem [{\citenamefont {Liu}\ \emph {et~al.}(2012)\citenamefont {Liu},
  \citenamefont {Pai}, \citenamefont {Li}, \citenamefont {Tseng}, \citenamefont
  {Ralph},\ and\ \citenamefont {Buhrman}}]{LiuS2012}%
  \BibitemOpen
  \bibfield  {author} {\bibinfo {author} {\bibfnamefont {L.}~\bibnamefont
  {Liu}}, \bibinfo {author} {\bibfnamefont {C.-F.}\ \bibnamefont {Pai}},
  \bibinfo {author} {\bibfnamefont {Y.}~\bibnamefont {Li}}, \bibinfo {author}
  {\bibfnamefont {H.}~\bibnamefont {Tseng}}, \bibinfo {author} {\bibfnamefont
  {D.}~\bibnamefont {Ralph}}, \ and\ \bibinfo {author} {\bibfnamefont
  {R.}~\bibnamefont {Buhrman}},\ }\href@noop {} {\bibfield  {journal} {\bibinfo
   {journal} {Science}\ }\textbf {\bibinfo {volume} {336}},\ \bibinfo {pages}
  {555} (\bibinfo {year} {2012})}\BibitemShut {NoStop}%
\bibitem [{\citenamefont {Garello}\ \emph {et~al.}(2014)\citenamefont
  {Garello}, \citenamefont {Avci}, \citenamefont {Miron}, \citenamefont
  {Baumgartner}, \citenamefont {Ghosh}, \citenamefont {Auffret}, \citenamefont
  {Boulle}, \citenamefont {Gaudin},\ and\ \citenamefont
  {Gambardella}}]{GarelloAPL2014}%
  \BibitemOpen
  \bibfield  {author} {\bibinfo {author} {\bibfnamefont {K.}~\bibnamefont
  {Garello}}, \bibinfo {author} {\bibfnamefont {C.~O.}\ \bibnamefont {Avci}},
  \bibinfo {author} {\bibfnamefont {I.~M.}\ \bibnamefont {Miron}}, \bibinfo
  {author} {\bibfnamefont {M.}~\bibnamefont {Baumgartner}}, \bibinfo {author}
  {\bibfnamefont {A.}~\bibnamefont {Ghosh}}, \bibinfo {author} {\bibfnamefont
  {S.}~\bibnamefont {Auffret}}, \bibinfo {author} {\bibfnamefont
  {O.}~\bibnamefont {Boulle}}, \bibinfo {author} {\bibfnamefont
  {G.}~\bibnamefont {Gaudin}}, \ and\ \bibinfo {author} {\bibfnamefont
  {P.}~\bibnamefont {Gambardella}},\ }\href {\doibase
  http://dx.doi.org/10.1063/1.4902443} {\bibfield  {journal} {\bibinfo
  {journal} {Appl. Phys. Lett.}\ }\textbf {\bibinfo {volume} {105}},\ \bibinfo
  {eid} {212402} (\bibinfo {year} {2014})}\BibitemShut {NoStop}%
\bibitem [{\citenamefont {Liu}\ \emph {et~al.}(2011)\citenamefont {Liu},
  \citenamefont {Moriyama}, \citenamefont {Ralph},\ and\ \citenamefont
  {Buhrman}}]{LiuPRL2011}%
  \BibitemOpen
  \bibfield  {author} {\bibinfo {author} {\bibfnamefont {L.}~\bibnamefont
  {Liu}}, \bibinfo {author} {\bibfnamefont {T.}~\bibnamefont {Moriyama}},
  \bibinfo {author} {\bibfnamefont {D.}~\bibnamefont {Ralph}}, \ and\ \bibinfo
  {author} {\bibfnamefont {R.}~\bibnamefont {Buhrman}},\ }\href@noop {}
  {\bibfield  {journal} {\bibinfo  {journal} {Phys. Rev. Lett.}\ }\textbf
  {\bibinfo {volume} {106}},\ \bibinfo {pages} {036601} (\bibinfo {year}
  {2011})}\BibitemShut {NoStop}%
\bibitem [{\citenamefont {Cubukcu}\ \emph {et~al.}(2014)\citenamefont
  {Cubukcu}, \citenamefont {Boulle}, \citenamefont {Drouard}, \citenamefont
  {Garello}, \citenamefont {Avci}, \citenamefont {Miron}, \citenamefont
  {Langer}, \citenamefont {Ocker}, \citenamefont {Gambardella},\ and\
  \citenamefont {Gaudin}}]{CubukcuAPL2014}%
  \BibitemOpen
  \bibfield  {author} {\bibinfo {author} {\bibfnamefont {M.}~\bibnamefont
  {Cubukcu}}, \bibinfo {author} {\bibfnamefont {O.}~\bibnamefont {Boulle}},
  \bibinfo {author} {\bibfnamefont {M.}~\bibnamefont {Drouard}}, \bibinfo
  {author} {\bibfnamefont {K.}~\bibnamefont {Garello}}, \bibinfo {author}
  {\bibfnamefont {C.~O.}\ \bibnamefont {Avci}}, \bibinfo {author}
  {\bibfnamefont {I.~M.}\ \bibnamefont {Miron}}, \bibinfo {author}
  {\bibfnamefont {J.}~\bibnamefont {Langer}}, \bibinfo {author} {\bibfnamefont
  {B.}~\bibnamefont {Ocker}}, \bibinfo {author} {\bibfnamefont
  {P.}~\bibnamefont {Gambardella}}, \ and\ \bibinfo {author} {\bibfnamefont
  {G.}~\bibnamefont {Gaudin}},\ }\href@noop {} {\bibfield  {journal} {\bibinfo
  {journal} {Appl. Phys. Lett.}\ }\textbf {\bibinfo {volume} {104}},\ \bibinfo
  {pages} {042406} (\bibinfo {year} {2014})}\BibitemShut {NoStop}%
\bibitem [{\citenamefont {Demidov}\ \emph {et~al.}(2012)\citenamefont
  {Demidov}, \citenamefont {Urazhdin}, \citenamefont {Ulrichs}, \citenamefont
  {Tiberkevich}, \citenamefont {Slavin}, \citenamefont {Baither}, \citenamefont
  {Schmitz},\ and\ \citenamefont {Demokritov}}]{DemidovNM2012}%
  \BibitemOpen
  \bibfield  {author} {\bibinfo {author} {\bibfnamefont {V.~E.}\ \bibnamefont
  {Demidov}}, \bibinfo {author} {\bibfnamefont {S.}~\bibnamefont {Urazhdin}},
  \bibinfo {author} {\bibfnamefont {H.}~\bibnamefont {Ulrichs}}, \bibinfo
  {author} {\bibfnamefont {V.}~\bibnamefont {Tiberkevich}}, \bibinfo {author}
  {\bibfnamefont {A.}~\bibnamefont {Slavin}}, \bibinfo {author} {\bibfnamefont
  {D.}~\bibnamefont {Baither}}, \bibinfo {author} {\bibfnamefont
  {G.}~\bibnamefont {Schmitz}}, \ and\ \bibinfo {author} {\bibfnamefont
  {S.~O.}\ \bibnamefont {Demokritov}},\ }\href@noop {} {\bibfield  {journal}
  {\bibinfo  {journal} {Nat. Mater.}\ }\textbf {\bibinfo {volume} {11}},\
  \bibinfo {pages} {1028} (\bibinfo {year} {2012})}\BibitemShut {NoStop}%
\bibitem [{\citenamefont {Nakayama}\ \emph {et~al.}(2013)\citenamefont
  {Nakayama}, \citenamefont {Althammer}, \citenamefont {Chen}, \citenamefont
  {Uchida}, \citenamefont {Kajiwara}, \citenamefont {Kikuchi}, \citenamefont
  {Ohtani}, \citenamefont {Gepr{\"a}gs}, \citenamefont {Opel}, \citenamefont
  {Takahashi} \emph {et~al.}}]{NakayamaPRL2013}%
  \BibitemOpen
  \bibfield  {author} {\bibinfo {author} {\bibfnamefont {H.}~\bibnamefont
  {Nakayama}}, \bibinfo {author} {\bibfnamefont {M.}~\bibnamefont {Althammer}},
  \bibinfo {author} {\bibfnamefont {Y.-T.}\ \bibnamefont {Chen}}, \bibinfo
  {author} {\bibfnamefont {K.}~\bibnamefont {Uchida}}, \bibinfo {author}
  {\bibfnamefont {Y.}~\bibnamefont {Kajiwara}}, \bibinfo {author}
  {\bibfnamefont {D.}~\bibnamefont {Kikuchi}}, \bibinfo {author} {\bibfnamefont
  {T.}~\bibnamefont {Ohtani}}, \bibinfo {author} {\bibfnamefont
  {S.}~\bibnamefont {Gepr{\"a}gs}}, \bibinfo {author} {\bibfnamefont
  {M.}~\bibnamefont {Opel}}, \bibinfo {author} {\bibfnamefont {S.}~\bibnamefont
  {Takahashi}},  \emph {et~al.},\ }\href@noop {} {\bibfield  {journal}
  {\bibinfo  {journal} {Phys. Rev. Lett.}\ }\textbf {\bibinfo {volume} {110}},\
  \bibinfo {pages} {206601} (\bibinfo {year} {2013})}\BibitemShut {NoStop}%
\bibitem [{\citenamefont {Hahn}\ \emph {et~al.}(2013)\citenamefont {Hahn},
  \citenamefont {De~Loubens}, \citenamefont {Klein}, \citenamefont {Viret},
  \citenamefont {Naletov},\ and\ \citenamefont {Youssef}}]{Hahn2013PRB}%
  \BibitemOpen
  \bibfield  {author} {\bibinfo {author} {\bibfnamefont {C.}~\bibnamefont
  {Hahn}}, \bibinfo {author} {\bibfnamefont {G.}~\bibnamefont {De~Loubens}},
  \bibinfo {author} {\bibfnamefont {O.}~\bibnamefont {Klein}}, \bibinfo
  {author} {\bibfnamefont {M.}~\bibnamefont {Viret}}, \bibinfo {author}
  {\bibfnamefont {V.~V.}\ \bibnamefont {Naletov}}, \ and\ \bibinfo {author}
  {\bibfnamefont {J.~B.}\ \bibnamefont {Youssef}},\ }\href@noop {} {\bibfield
  {journal} {\bibinfo  {journal} {Phys. Rev. B}\ }\textbf {\bibinfo {volume}
  {87}},\ \bibinfo {pages} {174417} (\bibinfo {year} {2013})}\BibitemShut
  {NoStop}%
\bibitem [{\citenamefont {Althammer}\ \emph {et~al.}(2013)\citenamefont
  {Althammer}, \citenamefont {Meyer}, \citenamefont {Nakayama}, \citenamefont
  {Schreier}, \citenamefont {Altmannshofer}, \citenamefont {Weiler},
  \citenamefont {Huebl}, \citenamefont {Gepr{\"a}gs}, \citenamefont {Opel},
  \citenamefont {Gross} \emph {et~al.}}]{althammer2013PRB}%
  \BibitemOpen
  \bibfield  {author} {\bibinfo {author} {\bibfnamefont {M.}~\bibnamefont
  {Althammer}}, \bibinfo {author} {\bibfnamefont {S.}~\bibnamefont {Meyer}},
  \bibinfo {author} {\bibfnamefont {H.}~\bibnamefont {Nakayama}}, \bibinfo
  {author} {\bibfnamefont {M.}~\bibnamefont {Schreier}}, \bibinfo {author}
  {\bibfnamefont {S.}~\bibnamefont {Altmannshofer}}, \bibinfo {author}
  {\bibfnamefont {M.}~\bibnamefont {Weiler}}, \bibinfo {author} {\bibfnamefont
  {H.}~\bibnamefont {Huebl}}, \bibinfo {author} {\bibfnamefont
  {S.}~\bibnamefont {Gepr{\"a}gs}}, \bibinfo {author} {\bibfnamefont
  {M.}~\bibnamefont {Opel}}, \bibinfo {author} {\bibfnamefont {R.}~\bibnamefont
  {Gross}},  \emph {et~al.},\ }\href@noop {} {\bibfield  {journal} {\bibinfo
  {journal} {Phys. Rev. B}\ }\textbf {\bibinfo {volume} {87}},\ \bibinfo
  {pages} {224401} (\bibinfo {year} {2013})}\BibitemShut {NoStop}%
\bibitem [{\citenamefont {Vlietstra}\ \emph {et~al.}(2014)\citenamefont
  {Vlietstra}, \citenamefont {Shan}, \citenamefont {van Wees}, \citenamefont
  {Isasa}, \citenamefont {Casanova},\ and\ \citenamefont
  {Youssef}}]{VlietstraPRB2014}%
  \BibitemOpen
  \bibfield  {author} {\bibinfo {author} {\bibfnamefont {N.}~\bibnamefont
  {Vlietstra}}, \bibinfo {author} {\bibfnamefont {J.}~\bibnamefont {Shan}},
  \bibinfo {author} {\bibfnamefont {B.}~\bibnamefont {van Wees}}, \bibinfo
  {author} {\bibfnamefont {M.}~\bibnamefont {Isasa}}, \bibinfo {author}
  {\bibfnamefont {F.}~\bibnamefont {Casanova}}, \ and\ \bibinfo {author}
  {\bibfnamefont {J.~B.}\ \bibnamefont {Youssef}},\ }\href@noop {} {\bibfield
  {journal} {\bibinfo  {journal} {Phys. Rev. B}\ }\textbf {\bibinfo {volume}
  {90}},\ \bibinfo {pages} {174436} (\bibinfo {year} {2014})}\BibitemShut
  {NoStop}%
\bibitem [{\citenamefont {V{\'e}lez}\ \emph {et~al.}(2015)\citenamefont
  {V{\'e}lez}, \citenamefont {Golovach}, \citenamefont {Isasa}, \citenamefont
  {Bedoya-Pinto}, \citenamefont {Sagasta}, \citenamefont {Pietrobon},
  \citenamefont {Hueso}, \citenamefont {Bergeret},\ and\ \citenamefont
  {Casanova}}]{VelezArx2015}%
  \BibitemOpen
  \bibfield  {author} {\bibinfo {author} {\bibfnamefont {S.}~\bibnamefont
  {V{\'e}lez}}, \bibinfo {author} {\bibfnamefont {V.~N.}\ \bibnamefont
  {Golovach}}, \bibinfo {author} {\bibfnamefont {M.}~\bibnamefont {Isasa}},
  \bibinfo {author} {\bibfnamefont {A.}~\bibnamefont {Bedoya-Pinto}}, \bibinfo
  {author} {\bibfnamefont {E.}~\bibnamefont {Sagasta}}, \bibinfo {author}
  {\bibfnamefont {L.}~\bibnamefont {Pietrobon}}, \bibinfo {author}
  {\bibfnamefont {L.~E.}\ \bibnamefont {Hueso}}, \bibinfo {author}
  {\bibfnamefont {F.~S.}\ \bibnamefont {Bergeret}}, \ and\ \bibinfo {author}
  {\bibfnamefont {F.}~\bibnamefont {Casanova}},\ }\href@noop {} {\bibfield
  {journal} {\bibinfo  {journal} {arXiv preprint arXiv:1502.04624}\ } (\bibinfo
  {year} {2015})}\BibitemShut {NoStop}%
\bibitem [{\citenamefont {Avci}\ \emph {et~al.}(2015)\citenamefont {Avci},
  \citenamefont {Garello}, \citenamefont {Ghosh}, \citenamefont {Gabureac},
  \citenamefont {Alvarado},\ and\ \citenamefont {Gambardella}}]{AvciNP2015}%
  \BibitemOpen
  \bibfield  {author} {\bibinfo {author} {\bibfnamefont {C.~O.}\ \bibnamefont
  {Avci}}, \bibinfo {author} {\bibfnamefont {K.}~\bibnamefont {Garello}},
  \bibinfo {author} {\bibfnamefont {A.}~\bibnamefont {Ghosh}}, \bibinfo
  {author} {\bibfnamefont {M.}~\bibnamefont {Gabureac}}, \bibinfo {author}
  {\bibfnamefont {S.~F.}\ \bibnamefont {Alvarado}}, \ and\ \bibinfo {author}
  {\bibfnamefont {P.}~\bibnamefont {Gambardella}},\ }\href@noop {} {\bibfield
  {journal} {\bibinfo  {journal} {Nat. Phys.}\ }\textbf {\bibinfo {volume}
  {11}},\ \bibinfo {pages} {570–575} (\bibinfo {year} {2015})}\BibitemShut
  {NoStop}%
\bibitem [{\citenamefont {Olejn{\'\i}k}\ \emph {et~al.}(2015)\citenamefont
  {Olejn{\'\i}k}, \citenamefont {Nov{\'a}k}, \citenamefont {Wunderlich},\ and\
  \citenamefont {Jungwirth}}]{OlejnikPRB2015}%
  \BibitemOpen
  \bibfield  {author} {\bibinfo {author} {\bibfnamefont {K.}~\bibnamefont
  {Olejn{\'\i}k}}, \bibinfo {author} {\bibfnamefont {V.}~\bibnamefont
  {Nov{\'a}k}}, \bibinfo {author} {\bibfnamefont {J.}~\bibnamefont
  {Wunderlich}}, \ and\ \bibinfo {author} {\bibfnamefont {T.}~\bibnamefont
  {Jungwirth}},\ }\href@noop {} {\bibfield  {journal} {\bibinfo  {journal}
  {Phys. Rev. B}\ }\textbf {\bibinfo {volume} {91}},\ \bibinfo {pages} {180402}
  (\bibinfo {year} {2015})}\BibitemShut {NoStop}%
\bibitem [{\citenamefont {Miao}\ \emph {et~al.}(2014)\citenamefont {Miao},
  \citenamefont {Huang}, \citenamefont {Qu},\ and\ \citenamefont
  {Chien}}]{Miao2014PRL}%
  \BibitemOpen
  \bibfield  {author} {\bibinfo {author} {\bibfnamefont {B.}~\bibnamefont
  {Miao}}, \bibinfo {author} {\bibfnamefont {S.}~\bibnamefont {Huang}},
  \bibinfo {author} {\bibfnamefont {D.}~\bibnamefont {Qu}}, \ and\ \bibinfo
  {author} {\bibfnamefont {C.}~\bibnamefont {Chien}},\ }\href@noop {}
  {\bibfield  {journal} {\bibinfo  {journal} {Phys. Rev. Lett.}\ }\textbf
  {\bibinfo {volume} {112}},\ \bibinfo {pages} {236601} (\bibinfo {year}
  {2014})}\BibitemShut {NoStop}%
\bibitem [{\citenamefont {Kobs}\ \emph {et~al.}(2011)\citenamefont {Kobs},
  \citenamefont {He{\ss}e}, \citenamefont {Kreuzpaintner}, \citenamefont
  {Winkler}, \citenamefont {Lott}, \citenamefont {Weinberger}, \citenamefont
  {Schreyer},\ and\ \citenamefont {Oepen}}]{KobsPRL2011}%
  \BibitemOpen
  \bibfield  {author} {\bibinfo {author} {\bibfnamefont {A.}~\bibnamefont
  {Kobs}}, \bibinfo {author} {\bibfnamefont {S.}~\bibnamefont {He{\ss}e}},
  \bibinfo {author} {\bibfnamefont {W.}~\bibnamefont {Kreuzpaintner}}, \bibinfo
  {author} {\bibfnamefont {G.}~\bibnamefont {Winkler}}, \bibinfo {author}
  {\bibfnamefont {D.}~\bibnamefont {Lott}}, \bibinfo {author} {\bibfnamefont
  {P.}~\bibnamefont {Weinberger}}, \bibinfo {author} {\bibfnamefont
  {A.}~\bibnamefont {Schreyer}}, \ and\ \bibinfo {author} {\bibfnamefont
  {H.}~\bibnamefont {Oepen}},\ }\href@noop {} {\bibfield  {journal} {\bibinfo
  {journal} {Phys. Rev. Lett.}\ }\textbf {\bibinfo {volume} {106}},\ \bibinfo
  {pages} {217207} (\bibinfo {year} {2011})}\BibitemShut {NoStop}%
\bibitem [{\citenamefont {Kobs}, \citenamefont {Frauen},\ and\ \citenamefont
  {Oepen}(2014)}]{KobsPRB2014}%
  \BibitemOpen
  \bibfield  {author} {\bibinfo {author} {\bibfnamefont {A.}~\bibnamefont
  {Kobs}}, \bibinfo {author} {\bibfnamefont {A.}~\bibnamefont {Frauen}}, \ and\
  \bibinfo {author} {\bibfnamefont {H.}~\bibnamefont {Oepen}},\ }\href@noop {}
  {\bibfield  {journal} {\bibinfo  {journal} {Phys. Rev. B}\ }\textbf {\bibinfo
  {volume} {90}},\ \bibinfo {pages} {016401} (\bibinfo {year}
  {2014})}\BibitemShut {NoStop}%
\bibitem [{\citenamefont {Lu}\ \emph {et~al.}(2013)\citenamefont {Lu},
  \citenamefont {Cai}, \citenamefont {Huang}, \citenamefont {Qu}, \citenamefont
  {Miao},\ and\ \citenamefont {Chien}}]{LuPRB2013}%
  \BibitemOpen
  \bibfield  {author} {\bibinfo {author} {\bibfnamefont {Y.}~\bibnamefont
  {Lu}}, \bibinfo {author} {\bibfnamefont {J.}~\bibnamefont {Cai}}, \bibinfo
  {author} {\bibfnamefont {S.}~\bibnamefont {Huang}}, \bibinfo {author}
  {\bibfnamefont {D.}~\bibnamefont {Qu}}, \bibinfo {author} {\bibfnamefont
  {B.}~\bibnamefont {Miao}}, \ and\ \bibinfo {author} {\bibfnamefont
  {C.}~\bibnamefont {Chien}},\ }\href@noop {} {\bibfield  {journal} {\bibinfo
  {journal} {Phys. Rev. B}\ }\textbf {\bibinfo {volume} {87}},\ \bibinfo
  {pages} {220409} (\bibinfo {year} {2013})}\BibitemShut {NoStop}%
\bibitem [{\citenamefont {Kim}\ \emph {et~al.}(2015)\citenamefont {Kim},
  \citenamefont {Sheng}, \citenamefont {Takahashi}, \citenamefont {Mitani},\
  and\ \citenamefont {Hayashi}}]{KimArx2015}%
  \BibitemOpen
  \bibfield  {author} {\bibinfo {author} {\bibfnamefont {J.}~\bibnamefont
  {Kim}}, \bibinfo {author} {\bibfnamefont {P.}~\bibnamefont {Sheng}}, \bibinfo
  {author} {\bibfnamefont {S.}~\bibnamefont {Takahashi}}, \bibinfo {author}
  {\bibfnamefont {S.}~\bibnamefont {Mitani}}, \ and\ \bibinfo {author}
  {\bibfnamefont {M.}~\bibnamefont {Hayashi}},\ }\href@noop {} {\bibfield
  {journal} {\bibinfo  {journal} {arXiv:1503.08903}\ } (\bibinfo {year}
  {2015})}\BibitemShut {NoStop}%
\bibitem [{\citenamefont {Cho}\ \emph {et~al.}(2015)\citenamefont {Cho},
  \citenamefont {Baek}, \citenamefont {Lee}, \citenamefont {Jo},\ and\
  \citenamefont {Park}}]{ChoSR2015}%
  \BibitemOpen
  \bibfield  {author} {\bibinfo {author} {\bibfnamefont {S.}~\bibnamefont
  {Cho}}, \bibinfo {author} {\bibfnamefont {S.-h.~C.}\ \bibnamefont {Baek}},
  \bibinfo {author} {\bibfnamefont {K.~D.}\ \bibnamefont {Lee}}, \bibinfo
  {author} {\bibfnamefont {Y.}~\bibnamefont {Jo}}, \ and\ \bibinfo {author}
  {\bibfnamefont {B.-G.}\ \bibnamefont {Park}},\ }\href@noop {} {\bibfield
  {journal} {\bibinfo  {journal} {Sci. Rep.}\ }\textbf {\bibinfo {volume} {5}}
  (\bibinfo {year} {2015})}\BibitemShut {NoStop}%
\bibitem [{\citenamefont {Smith}, \citenamefont {Janak},\ and\ \citenamefont
  {Adler}(1967)}]{Smith1967}%
  \BibitemOpen
  \bibfield  {author} {\bibinfo {author} {\bibfnamefont {A.~C.}\ \bibnamefont
  {Smith}}, \bibinfo {author} {\bibfnamefont {J.~F.}\ \bibnamefont {Janak}}, \
  and\ \bibinfo {author} {\bibfnamefont {R.~B.}\ \bibnamefont {Adler}},\
  }\href@noop {} {\emph {\bibinfo {title} {Electronic conduction in solids}}}\
  (\bibinfo  {publisher} {McGraw-Hill New York},\ \bibinfo {year}
  {1967})\BibitemShut {NoStop}%
\bibitem [{\citenamefont {Weiler}\ \emph {et~al.}(2013)\citenamefont {Weiler},
  \citenamefont {Althammer}, \citenamefont {Schreier}, \citenamefont {Lotze},
  \citenamefont {Pernpeintner}, \citenamefont {Meyer}, \citenamefont {Huebl},
  \citenamefont {Gross}, \citenamefont {Kamra}, \citenamefont {Xiao} \emph
  {et~al.}}]{WeilerPRL2013}%
  \BibitemOpen
  \bibfield  {author} {\bibinfo {author} {\bibfnamefont {M.}~\bibnamefont
  {Weiler}}, \bibinfo {author} {\bibfnamefont {M.}~\bibnamefont {Althammer}},
  \bibinfo {author} {\bibfnamefont {M.}~\bibnamefont {Schreier}}, \bibinfo
  {author} {\bibfnamefont {J.}~\bibnamefont {Lotze}}, \bibinfo {author}
  {\bibfnamefont {M.}~\bibnamefont {Pernpeintner}}, \bibinfo {author}
  {\bibfnamefont {S.}~\bibnamefont {Meyer}}, \bibinfo {author} {\bibfnamefont
  {H.}~\bibnamefont {Huebl}}, \bibinfo {author} {\bibfnamefont
  {R.}~\bibnamefont {Gross}}, \bibinfo {author} {\bibfnamefont
  {A.}~\bibnamefont {Kamra}}, \bibinfo {author} {\bibfnamefont
  {J.}~\bibnamefont {Xiao}},  \emph {et~al.},\ }\href@noop {} {\bibfield
  {journal} {\bibinfo  {journal} {Phys. Rev. Lett.}\ }\textbf {\bibinfo
  {volume} {111}},\ \bibinfo {pages} {176601} (\bibinfo {year}
  {2013})}\BibitemShut {NoStop}%
\bibitem [{\citenamefont {Jungfleisch}\ \emph {et~al.}(2013)\citenamefont
  {Jungfleisch}, \citenamefont {Lauer}, \citenamefont {Neb}, \citenamefont
  {Chumak},\ and\ \citenamefont {Hillebrands}}]{JungfleischAPL2013}%
  \BibitemOpen
  \bibfield  {author} {\bibinfo {author} {\bibfnamefont {M.}~\bibnamefont
  {Jungfleisch}}, \bibinfo {author} {\bibfnamefont {V.}~\bibnamefont {Lauer}},
  \bibinfo {author} {\bibfnamefont {R.}~\bibnamefont {Neb}}, \bibinfo {author}
  {\bibfnamefont {A.}~\bibnamefont {Chumak}}, \ and\ \bibinfo {author}
  {\bibfnamefont {B.}~\bibnamefont {Hillebrands}},\ }\href@noop {} {\bibfield
  {journal} {\bibinfo  {journal} {Applied Physics Letters}\ }\textbf {\bibinfo
  {volume} {103}},\ \bibinfo {pages} {022411} (\bibinfo {year}
  {2013})}\BibitemShut {NoStop}%
\bibitem [{\citenamefont {Isasa}\ \emph {et~al.}(2014)\citenamefont {Isasa},
  \citenamefont {Bedoya-Pinto}, \citenamefont {V{\'e}lez}, \citenamefont
  {Golmar}, \citenamefont {S{\'a}nchez}, \citenamefont {Hueso}, \citenamefont
  {Fontcuberta},\ and\ \citenamefont {Casanova}}]{IsasaAPL2014}%
  \BibitemOpen
  \bibfield  {author} {\bibinfo {author} {\bibfnamefont {M.}~\bibnamefont
  {Isasa}}, \bibinfo {author} {\bibfnamefont {A.}~\bibnamefont {Bedoya-Pinto}},
  \bibinfo {author} {\bibfnamefont {S.}~\bibnamefont {V{\'e}lez}}, \bibinfo
  {author} {\bibfnamefont {F.}~\bibnamefont {Golmar}}, \bibinfo {author}
  {\bibfnamefont {F.}~\bibnamefont {S{\'a}nchez}}, \bibinfo {author}
  {\bibfnamefont {L.~E.}\ \bibnamefont {Hueso}}, \bibinfo {author}
  {\bibfnamefont {J.}~\bibnamefont {Fontcuberta}}, \ and\ \bibinfo {author}
  {\bibfnamefont {F.}~\bibnamefont {Casanova}},\ }\href@noop {} {\bibfield
  {journal} {\bibinfo  {journal} {Appl. Phys. Lett.}\ }\textbf {\bibinfo
  {volume} {105}},\ \bibinfo {pages} {142402} (\bibinfo {year}
  {2014})}\BibitemShut {NoStop}%
\bibitem [{\citenamefont {Grigoryan}\ \emph {et~al.}(2014)\citenamefont
  {Grigoryan}, \citenamefont {Guo}, \citenamefont {Bauer}, \citenamefont {Xiao}
  \emph {et~al.}}]{GrigoryanPRB2014}%
  \BibitemOpen
  \bibfield  {author} {\bibinfo {author} {\bibfnamefont {V.~L.}\ \bibnamefont
  {Grigoryan}}, \bibinfo {author} {\bibfnamefont {W.}~\bibnamefont {Guo}},
  \bibinfo {author} {\bibfnamefont {G.~E.}\ \bibnamefont {Bauer}}, \bibinfo
  {author} {\bibfnamefont {J.}~\bibnamefont {Xiao}},  \emph {et~al.},\
  }\href@noop {} {\bibfield  {journal} {\bibinfo  {journal} {Phys. Rev. B}\
  }\textbf {\bibinfo {volume} {90}},\ \bibinfo {pages} {161412} (\bibinfo
  {year} {2014})}\BibitemShut {NoStop}%
\bibitem [{\citenamefont {Avci}\ \emph
  {et~al.}(2014{\natexlab{a}})\citenamefont {Avci}, \citenamefont {Garello},
  \citenamefont {Gabureac}, \citenamefont {Ghosh}, \citenamefont {Fuhrer},
  \citenamefont {Alvarado},\ and\ \citenamefont {Gambardella}}]{AvciPRB2014b}%
  \BibitemOpen
  \bibfield  {author} {\bibinfo {author} {\bibfnamefont {C.~O.}\ \bibnamefont
  {Avci}}, \bibinfo {author} {\bibfnamefont {K.}~\bibnamefont {Garello}},
  \bibinfo {author} {\bibfnamefont {M.}~\bibnamefont {Gabureac}}, \bibinfo
  {author} {\bibfnamefont {A.}~\bibnamefont {Ghosh}}, \bibinfo {author}
  {\bibfnamefont {A.}~\bibnamefont {Fuhrer}}, \bibinfo {author} {\bibfnamefont
  {S.~F.}\ \bibnamefont {Alvarado}}, \ and\ \bibinfo {author} {\bibfnamefont
  {P.}~\bibnamefont {Gambardella}},\ }\href {\doibase
  10.1103/PhysRevB.90.224427} {\bibfield  {journal} {\bibinfo  {journal} {Phys.
  Rev. B}\ }\textbf {\bibinfo {volume} {90}},\ \bibinfo {pages} {224427}
  (\bibinfo {year} {2014}{\natexlab{a}})}\BibitemShut {NoStop}%
\bibitem [{\citenamefont {Pai}\ \emph {et~al.}(2012)\citenamefont {Pai},
  \citenamefont {Liu}, \citenamefont {Li}, \citenamefont {Tseng}, \citenamefont
  {Ralph},\ and\ \citenamefont {Buhrman}}]{PaiAPL2012}%
  \BibitemOpen
  \bibfield  {author} {\bibinfo {author} {\bibfnamefont {C.-F.}\ \bibnamefont
  {Pai}}, \bibinfo {author} {\bibfnamefont {L.}~\bibnamefont {Liu}}, \bibinfo
  {author} {\bibfnamefont {Y.}~\bibnamefont {Li}}, \bibinfo {author}
  {\bibfnamefont {H.}~\bibnamefont {Tseng}}, \bibinfo {author} {\bibfnamefont
  {D.}~\bibnamefont {Ralph}}, \ and\ \bibinfo {author} {\bibfnamefont
  {R.}~\bibnamefont {Buhrman}},\ }\href@noop {} {\bibfield  {journal} {\bibinfo
   {journal} {Appl. Phys. Lett.}\ }\textbf {\bibinfo {volume} {101}},\ \bibinfo
  {pages} {122404} (\bibinfo {year} {2012})}\BibitemShut {NoStop}%
\bibitem [{\citenamefont {Avci}\ \emph
  {et~al.}(2014{\natexlab{b}})\citenamefont {Avci}, \citenamefont {Garello},
  \citenamefont {Nistor}, \citenamefont {Godey}, \citenamefont {Ballesteros},
  \citenamefont {Mugarza}, \citenamefont {Barla}, \citenamefont {Valvidares},
  \citenamefont {Pellegrin}, \citenamefont {Ghosh}, \citenamefont {Miron},
  \citenamefont {Boulle}, \citenamefont {Auffret}, \citenamefont {Gaudin},\
  and\ \citenamefont {Gambardella}}]{AvciPRB2014}%
  \BibitemOpen
  \bibfield  {author} {\bibinfo {author} {\bibfnamefont {C.~O.}\ \bibnamefont
  {Avci}}, \bibinfo {author} {\bibfnamefont {K.}~\bibnamefont {Garello}},
  \bibinfo {author} {\bibfnamefont {C.}~\bibnamefont {Nistor}}, \bibinfo
  {author} {\bibfnamefont {S.}~\bibnamefont {Godey}}, \bibinfo {author}
  {\bibfnamefont {B.}~\bibnamefont {Ballesteros}}, \bibinfo {author}
  {\bibfnamefont {A.}~\bibnamefont {Mugarza}}, \bibinfo {author} {\bibfnamefont
  {A.}~\bibnamefont {Barla}}, \bibinfo {author} {\bibfnamefont
  {M.}~\bibnamefont {Valvidares}}, \bibinfo {author} {\bibfnamefont
  {E.}~\bibnamefont {Pellegrin}}, \bibinfo {author} {\bibfnamefont
  {A.}~\bibnamefont {Ghosh}}, \bibinfo {author} {\bibfnamefont {I.~M.}\
  \bibnamefont {Miron}}, \bibinfo {author} {\bibfnamefont {O.}~\bibnamefont
  {Boulle}}, \bibinfo {author} {\bibfnamefont {S.}~\bibnamefont {Auffret}},
  \bibinfo {author} {\bibfnamefont {G.}~\bibnamefont {Gaudin}}, \ and\ \bibinfo
  {author} {\bibfnamefont {P.}~\bibnamefont {Gambardella}},\ }\href@noop {}
  {\bibfield  {journal} {\bibinfo  {journal} {Phys. Rev. B}\ }\textbf {\bibinfo
  {volume} {89}},\ \bibinfo {pages} {214419} (\bibinfo {year}
  {2014}{\natexlab{b}})}\BibitemShut {NoStop}%
\bibitem [{\citenamefont {Nguyen}, \citenamefont {Pratt~Jr},\ and\
  \citenamefont {Bass}(2014)}]{bass2014jmmm}%
  \BibitemOpen
  \bibfield  {author} {\bibinfo {author} {\bibfnamefont {H.}~\bibnamefont
  {Nguyen}}, \bibinfo {author} {\bibfnamefont {W.}~\bibnamefont {Pratt~Jr}}, \
  and\ \bibinfo {author} {\bibfnamefont {J.}~\bibnamefont {Bass}},\ }\href@noop
  {} {\bibfield  {journal} {\bibinfo  {journal} {J. Magn. Magn. Mat.}\ }\textbf
  {\bibinfo {volume} {361}},\ \bibinfo {pages} {30} (\bibinfo {year}
  {2014})}\BibitemShut {NoStop}%
\bibitem [{\citenamefont {Vlietstra}\ \emph {et~al.}(2013)\citenamefont
  {Vlietstra}, \citenamefont {Shan}, \citenamefont {Castel}, \citenamefont {van
  Wees},\ and\ \citenamefont {Youssef}}]{vlietstra2013PRB}%
  \BibitemOpen
  \bibfield  {author} {\bibinfo {author} {\bibfnamefont {N.}~\bibnamefont
  {Vlietstra}}, \bibinfo {author} {\bibfnamefont {J.}~\bibnamefont {Shan}},
  \bibinfo {author} {\bibfnamefont {V.}~\bibnamefont {Castel}}, \bibinfo
  {author} {\bibfnamefont {B.}~\bibnamefont {van Wees}}, \ and\ \bibinfo
  {author} {\bibfnamefont {J.~B.}\ \bibnamefont {Youssef}},\ }\href@noop {}
  {\bibfield  {journal} {\bibinfo  {journal} {Phys. Rev. B}\ }\textbf {\bibinfo
  {volume} {87}},\ \bibinfo {pages} {184421} (\bibinfo {year}
  {2013})}\BibitemShut {NoStop}%
\bibitem [{\citenamefont {Schreier}\ \emph {et~al.}(2013)\citenamefont
  {Schreier}, \citenamefont {Roschewsky}, \citenamefont {Dobler}, \citenamefont
  {Meyer}, \citenamefont {Huebl}, \citenamefont {Gross},\ and\ \citenamefont
  {Goennenwein}}]{SchreierAPL2013}%
  \BibitemOpen
  \bibfield  {author} {\bibinfo {author} {\bibfnamefont {M.}~\bibnamefont
  {Schreier}}, \bibinfo {author} {\bibfnamefont {N.}~\bibnamefont
  {Roschewsky}}, \bibinfo {author} {\bibfnamefont {E.}~\bibnamefont {Dobler}},
  \bibinfo {author} {\bibfnamefont {S.}~\bibnamefont {Meyer}}, \bibinfo
  {author} {\bibfnamefont {H.}~\bibnamefont {Huebl}}, \bibinfo {author}
  {\bibfnamefont {R.}~\bibnamefont {Gross}}, \ and\ \bibinfo {author}
  {\bibfnamefont {S.~T.}\ \bibnamefont {Goennenwein}},\ }\href@noop {}
  {\bibfield  {journal} {\bibinfo  {journal} {Appl. Phys. Lett.}\ }\textbf
  {\bibinfo {volume} {103}},\ \bibinfo {pages} {242404} (\bibinfo {year}
  {2013})}\BibitemShut {NoStop}%
\end{thebibliography}

\end{document}